\documentclass[aps,pre,twocolumn,longbibliography,preprintnumbers,amsmath,amssymb,showkeys,reprint]{revtex4-1}
\pdfoutput=1

\usepackage{graphicx}
\usepackage{dcolumn}
\usepackage{bm}
\usepackage{upgreek}
\usepackage{array}
\usepackage{booktabs}
\usepackage{microtype}
\usepackage{cmap}
\usepackage{geometry}
\usepackage[dvipsnames]{xcolor}

\usepackage[pdftex]{hyperref}
\hypersetup{citecolor={blue}, urlcolor={violet},  linkcolor={red}, colorlinks=true}

\begin{document}

\author{Ivan N. Terterov}
\affiliation{School of Physics and Engineering, ITMO University, St. Petersburg, Russia}
\email{ivan.terterov@metalab.ifmo.ru}

\author{Sergei V. Koniakhin}
\affiliation{Center for Theoretical Physics of Complex Systems, Institute for Basic Science (IBS), Daejeon 34126, Republic of Korea}

\author{Alexey A. Bogdanov }
\affiliation{St.-Petersburg Clinical Scientific and Practical Center of Specialized Types of Medical Care (Oncological), St. Petersburg, Russia}

\title{Specific and non-specific effects of sodium and potassium ions on the interactions between model charged groups of proteins}





\begin{abstract}
Potassium and sodium ions are crucial for many physiological processes in living systems and play different roles when interacting with proteins and enzymes. Intracellular concentration of potassium is always maintained higher than that of sodium, which provides a suitable environment for biochemical machinery. These cations also possess different properties in physico-chemical Hofmeister phenomena. It is now accepted that the main physical reason for these ion-specific effects is due to formation of ion pairs. The greater ability of sodium over potassium to destabilize protein solutions was previously rationalized by sodium stronger pairing with carboxylates, which are the main anionic moiety of proteins. While ion pairing of cations with carboxylates was studied in detail previously, understanding of the molecular mechanisms of cation-specific mediation of protein-protein interactions is still lacking. In this work with the use of molecular dynamics we studied the effect of sodium and potassium on interactions between model compounds that bear typical positively and negatively charged groups of proteins, namely, methylammonium and acetate molecules. We found only a weak difference in the contacts of charged groups depending on the cation type present in the solution. Our results suggest that a strength of cation-carboxylate binding is not critical, but structural features may be more important.

\end{abstract}
\maketitle


\section{Introduction}
Potassium and sodium are the most abundant monovalent metal ions in living systems. These ions are very similar, but play different roles in many physiological processes, such as transmembrane transport, electroexcitation of cells, enzymatic activity and many others. The concentration of potassium ions inside living cells is higher than that of sodium, while the concentration ratio in the extracellular medium is the opposite \cite[]{albe_2002_book}. Na$^+$ goes adjacent to K$^+$ in the Hofmeister series of cations, which ranges them by the ability to promote protein precipitation \cite[]{collins2004ions}. It indicates that sodium is more effective in the protein solution destabilization, leading to so-called "salting-out" effect. Interestingly, same series of ions were observed when ranking the strength of physico-chemical ion-specific effects like surface-tension, properties of polymers and others \cite[]{lo2012hofmeister}. While the difference of Na$^+$ and K$^+$ specific effects on protein stability is not dramatic compared to the case of divalent cations or distinct anions \cite{lo2012hofmeister}, it is noticeable in several cases \cite[]{kherb2012role,kastelic2015protein} and is thought to be of grate significance for biological processes inside living cells \cite[]{vrbka2006quantification}.

It is now widely accepted that the main physical mechanism underlying Hofmeister effect is the specific pairing of ions with chemical groups on the protein surfaces \cite[]{lo2012hofmeister,van2016water,jungwirth2014beyond}. Major anionic group represented in protein structures is a carboxilate, which is a part of the aspartate and glutamate residues. Thus for sodium versus potassium problem ion pairs with carboxylate are of central importance, which are usually investigated in solutions of simple model molecules that contain this group (e.g. acetate salts solutions). Particularly, in potentiometric studies it was shown that  Na$^+$ and K$^+$ form weak ion pairs with caboxylate, which are slightly more stable for sodium \cite[]{daniele2008weak}. However, with the method mentioned above it is hard to decipher molecular details, particularly to separate direct contact ion pair (CIP) and solvent-shared ion pair (SIP) -- coordination mode, characterized by the presence of water molecule between cation and carboxylate \cite[]{marcus2006ion}. More specific data on CIP was obtained using X-ray absorption spectroscopy, which also indicates that sodium ions bind to a negatively charged carboxyl group stronger than potassium ions \cite[]{aziz2008cation}. Raman spectroscopy also is thought to be more sensitive for CIP rather than SIP \cite[]{fournier1998experimental}, however recent data indicate that it is difficult to resolve sodium carboxylate ion pairing with Raman spectra \cite[]{de2020binding}. Further molecular details were investigated with molecular dynamics (MD) for systems with proteins \cite[]{friedman2011ions,heyda2009ion,vrbka2006quantification} as well as for systems containing acetate ions in order to direct comparison with experimental thermodynamic data \cite[]{jagoda2007ion,hess2009cation,annapureddy2012molecular,ahlstrand2017computer}. These studies demonstrate that ion coordination around carboxylate sufficiently differs for sodium and potassium, and typically sodium binds strongly in accord with experimental results.

All these finding suggest that functional protein-protein interactions that are usually determined by charge-charge contacts may be affected differently by sodium and potassium ions. The main idea is that due to stronger binding, sodium will more effectively than potassium hinder anionic groups of protein and thus interfering with their functional contacts with cationic groups of other proteins \cite[]{vrbka2006quantification}. However, demonstration of a molecular details underlining this possible mechanism is still lacking. To address this issue by means of molecular dynamics we investigated specific effects that may be produced by Na$^+$ and K$^+$ on the charge-charge interaction between model chemical groups of proteins. For simplicity we considered binding of two model solutes that bear either a characteristic anionic or cationic group. Specifically, we calculated a potentials of mean force between carboxylate on acetate molecule and charged amino-group carried by methylammonium molecule, and investigated the influence exerted on this interactions either by Na$^+$ or K$^+$ ions presented in solution in different concentrations

\section{Methods}
To investigate the difference in Na$^+$ and K$^+$ effects on the interactions of protein charged groups, we utilized the approach inspired by the work Thomas and Elcock \cite{thomas2006direct}, in which it was demonstrated that it is possible to assess salt effects on the interaction between model solutes directly within unbiased molecular dynamics (MD) simulation. As in the original work \cite{thomas2006direct}, we used acetate and methylammonium molecules as model solutes. Acetate bears carboxylic group in anionic state, which may be found on ASP or GLU residues and at the C-terminus of proteins, while methylammonium contains positively charged amino-group, which is a part of LYS residue and a protein N-terminus. Both molecules also contain one hydrophobic methyl group. Force field parameters for model solutes were taken from OPLS-AA force field \cite{jorgensen1996development} (from GLU, LYS and LEU residues). It is known that standard force fields overestimate ion pairing \cite{hess2009cation,yoo2018new,kirby2019charge,tolmachev2019overbinding}, to take this possible source of artifacts into account we used two different sets of ion parameters for comparison. First force field (FF) we used for ions combines parameters by Aqvist \cite{aqvist1990ion} for Na$^+$, K$^+$ and Dang \cite{dang1995mechanism} for CL$^{-}$ (this parameter set hereafter are demoted as "Aqvist FF"). This combination of parameters is quite often used, however known to overestimate sodium binding to anions \cite{joung2008determination,hess2009cation}. Second ion force field we used was specifically developed to correctly reproduce experimental thermodynamic data of cation-acetate interactions based on Kirkwood-Buff solution theory \cite{hess2009cation,weerasinghe2003kirkwood} (hereafter it is denoted as "KB FF"). As a parameters in KB FF were developed with SPC/E water model \cite{berendsen1987missing} we used this water model for all simulations for universality.

We investigated systems, which contained one molecule of acetate and one molecule of methylammonium solvated in a periodic cubic box together with various number of ions that lead to different NaCl or KCl concentrations (0.15M, 0.3M, 0.5M , 1M , 2M), or without salts (i.e. acetate and methylammonium pair in pure water). In the original work Thomas and Elcock \cite{thomas2006direct} showed that for systems of 2.5x2.5x2.5 nm$^3$ box size indirect effects due to interactions with periodic images are negligible for 0.3M and higher salt concentrations. To ensure that there is no unwanted effects from the periodic images we used larger 3x3x3 nm$^3$ cubic system for concentrations 0.3M, 0.5M , 1M , 2M, which corresponds to systems with 5, 8, 16 and 32 anions (CL$^{-}$) and cation ions (either Na$^+$ or K$^+$), respectively. To minimize the influence of periodicity in case of salt free system and for physiological salt concentration 0.15M we utilized slightly different approach. In order to reduce effective energy of interactions between periodic images for these systems we used a larger cubic box (4.45x4.45x4.45 nm$^3$), moreover the distance between centers of charge of molecules was limited to be maximum 1.5 nm with flat-bottom potential between carbon atom of acetate carboxyl groups and nitrogen atom of methylammonium (force constant of 5000 $kJ/(mol\cdot nm^2$)). This potential equals zero for lower distances thus it does not disturb any mutual conformations of acetate and methylammonium, which are the most interesting for the present study. At the same time this potential restricts sampling of irrelevant distant configurations, which are also more affected by interactions with periodic images. Resulting systems for physiological salt concentration (0.15M) together with acetate-methylammonium pair contained 8 CL$^{-}$ ions and 8 cations (either Na$^+$ or K$^+$).

All simulations were conducted with GROMACS software package \cite{abraham2015gromacs} (version 2018.6) in NPT ensemble, constant pressure was maintained at 1 bar with Parrinello-Rahman barostat \cite{parrinello1981polymorphic}, temperature was controlled with Nose-Hoover thermostat at 300 K \cite{nose1984unified,hoover1985canonical}. After a 10 ns of equilibration data from a 1.5 $\mu$s long simulations were collected for systems with 0.3M, 0.5M , 1M , 2M salt concentrations, and 2 $\mu$s for 0.15M and salt free systems. Additionally, to test ion coordination around free crboxylate we  conducted 0.5 $\mu$s MD simulations of systems containing only one acetate molecule (without methylammonium) and either 1M NaCl or 1M KCl (16 cations and 15 CL$^{-}$ ions). For all considered system variants, two trajectories were obtained: one with Aqvist FF and the other with KB FF.

During long unbiased MD trajectories mutual acetate-methylammonium configurations were sampled sufficiently to obtain potential of mean force (PMF) between molecules. Along a selected intermolecular reaction coordinate ($\mathbf{r}$) the PMF ($W(\mathbf{r})$) may be calculated as follows \cite{thomas2006direct}
\begin{equation}
\label{eq:pmf}
W(\mathbf{r}) = -RT \ln \left[ P(\mathbf{r})/  P^{ig}(\mathbf{r}) \right],
\end{equation}
where $P(\mathbf{r})$ is a probability of finding the system in configurations with coordinate $\mathbf{r}$ during the unbiased MD, $P^{ig}(\mathbf{r})$ -- "ideal-gas" probability of the same states in the system where all species do not interact,  $R$ --  gas constant and $T$ -- temperature. If reaction coordinate is defined as the distance between selected atoms, the $P(\mathbf{r})$ is simply computed as a probability of configurations in MD trajectory for which the distance is in the short interval $[r, r + dr)$. Corresponding "ideal-gas" probability reflects the accessible volume for this interval in $\mathbf{r}$, which is a spherical layer  $P^{ig}(\mathbf{r}) \sim 4 \pi r^2 dr$. For the analysis, we considered three options for reaction coordinates between acetate and methylammonium molecules: (i) the distance between acetate carbon atom of carboxyl group ($COO^-$) and nitrogen atom of methylammonium, (ii) the distance from methylammonium nitrogen atom to a nearest oxygen atom of $COO^-$ group of acetate and (iii) the distance between carbon atoms of methyl groups of acetate and methylammonium. In case (ii) non-spherical accessible volume was taken into account when calculating $P^{ig}(\mathbf{r})$ \cite{hess2009cation}.

It should be noted that the PMFs obtained with the equation (\ref{eq:pmf}) is defined only up to an additive constant, which comes from the normalization of probabilities $P(r)$ and $P^{ig}(r)$. Thus such defined PMFs only give relative free energies for different distances. To quantitatively compare PMFs for systems with different salt concentrations, each obtained potential should be transfered to an absolute scale. We propose that computed PMFs at sufficiently large distance (where solvent structure effects may be neglected) should follow potential of continuum electrostatics. For salt free systems we used as reference a Coulomb potential with the relative electric permittivity constant for SPC/E water model $\varepsilon = 68$ \cite{vega2011simulating}. For systems with salt as a reference we used a screened Coulomb potential derived in the Debye-Huckel theory of electrolytes \cite{debye1954collected}
\begin{equation}
\label{eq:scr_coulomb}
U_{scr}(r) = \frac{\exp(-\kappa \cdot r)}{4\pi \varepsilon \varepsilon_0 r}, 
\end{equation}
where $\varepsilon$ and  $\varepsilon_0$ are electric permittivity constants of water (for SPC/E, $\varepsilon = 68$ \cite{vega2011simulating}) and vacuum, respectively; and $\kappa$ -- Debye parameter, which is defined as $\kappa = \sqrt{2000 N_a q_e^2 /(\varepsilon \varepsilon_0 k T)I}$, where $N_a$ and $k$ -- are Avagrdro and Boltzmann constants, $T$ -- temperature, $q_e$ -- elementary charge and $I$ -- ionic strength of the solution in mole/l units (M). In case of 1:1 electrolyte such as NaCl and KCl solutions the ionic strength ($I$) equals salt concentration, which we calculated for each system as an average value over MD trajectory. PMFs on the absolute scale were obtained after shifting $W(r)$ curves calculated with (\ref{eq:pmf}) so as to fit the corresponding reference potentials (\ref{eq:scr_coulomb}). It was found that PMF curves are well fitted to the reference potentials starting at distances of about $1.0$ - $1.2$ nm (see Results for details). 

With corrected PMFs the absolute association constant and free energy for acetate-methylammonium binding under the conditions of each simulated system were calculated as follows \cite{zhang2003studying}
\begin{eqnarray}
\label{eq:k_a}
K_a = \frac{1}{V_{1M}} \int_0^{r_c} dr\, 4\pi r^2 \exp\left[-W(r)/RT\right] , \nonumber\\
\Delta G = -RT \ln[K_a],
\end{eqnarray}
where the bound state is defined by the condition that the distance $r$ is lower than some threshold $r_c$, $V_{1M} = 1.66$ nm$^3$ -- is a normalization volume needed to establish $1M$ standard state.

\section{Results and discussion}
\subsection{Cation coordination around carboxylate}
\begin{figure*}[ht]
  \includegraphics[width=1.0\linewidth]{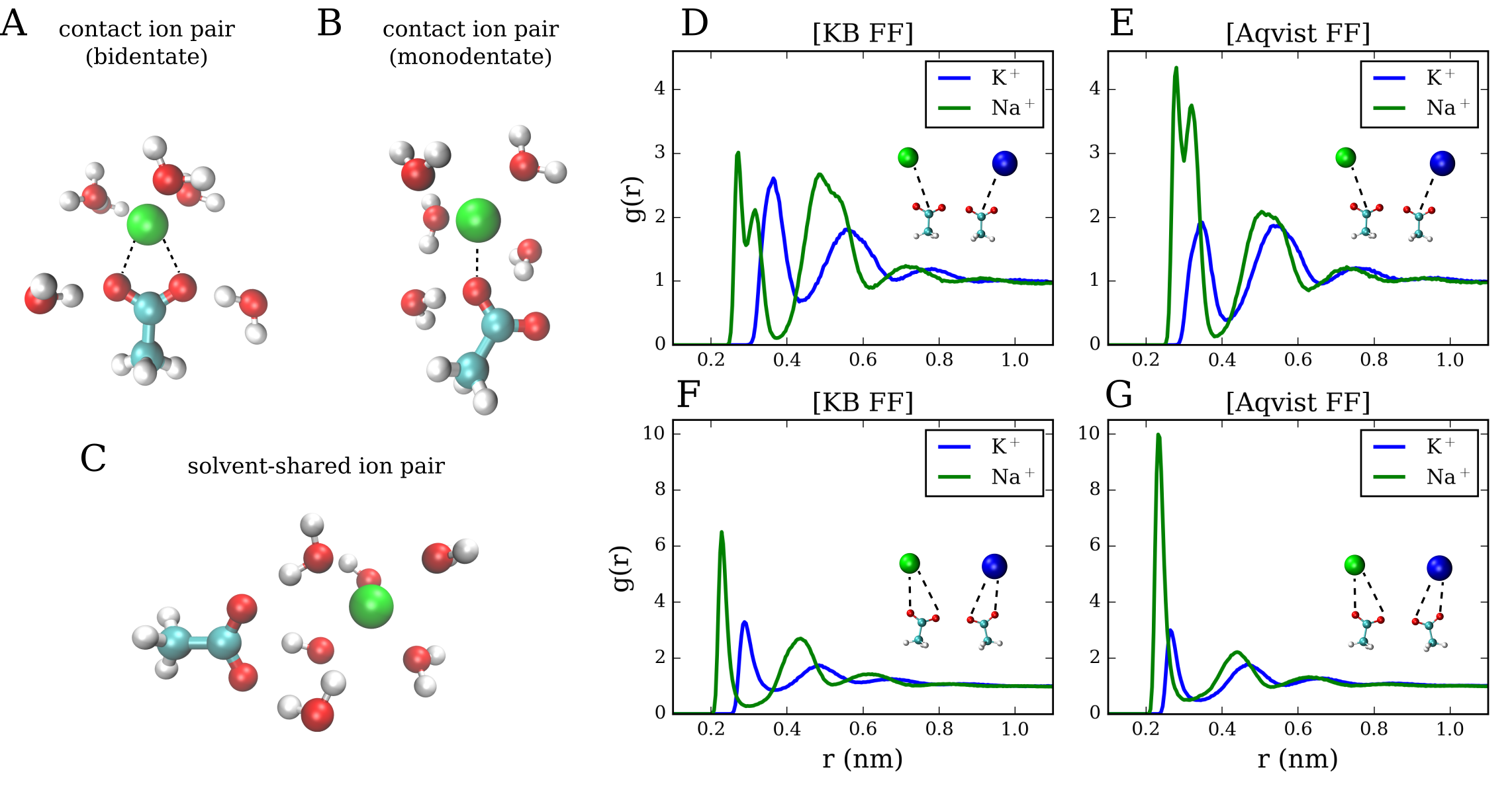}
  \caption{Cation coordination around carboxylate. (A-C) Snapshots of representative ion-acetate coordination structures: contact ion pair (CIP) in bidentate configuration (panel A), CIP in monodentate configuration (panel B) and solvent-shared ion pair (SIP, panel C); sodium, oxygen, carbon, and hydrogen atoms are shown in green, red, gray, and white colors. (D-E) Radial distribution functions (RDFs) between cations and carboxyl carbon atom calculated from simulations with parameters from KB FF (D) and Aqvist FF (E). (F-G) RDFs between cations and carboxyl oxygen atoms for KB FF (F) and Aqvist FF (G). Distances for corresponding RDFs are illustrated with dashed lines in the inserts (green sphere is Na$^+$ ion, blue -- K$^+$). RDFs for Na$^+$ are shown with green, for K$^+$ -- with blue lines.} 
  \label{fig:rdfs_one}
\end{figure*}

We firstly focus on the coordination of Na$^+$ and K$^+$ ions around carboxylate group, investigated within systems that contain only one acetate molecule in 1M salt solution. Radial distribution functions (RDFs) between cations and carboxyl oxygen atoms are shown in Figure \ref{fig:rdfs_one} F and G, and it is clearly seen that there is a first main peak of RDFs, which corresponds to the contact ion pair (CIP), followed by the next smaller peak, which corresponds to the solvent-shared ion pair (SIP) (examples of ion pair structures are presented in Figure \ref{fig:rdfs_one}A, B, C). Same overall picture is observed on RDFs between cations and carboxyl carbon (Figure \ref{fig:rdfs_one} D and E), except that in the sodium RDF first peak ($0.2<r<0.4$ nm) is split into two subpeaks for both used force fields, which is not observed for potassium. The nearest subpeak corresponds to the bidentate configuration of CIP, in which both carboxylate oxygen atoms are in direct contact with the ion (Figure \ref{fig:rdfs_one} A). Second subpeak indicate the monodentate configuration when only one oxygen is in direct contact with the ion (Figure \ref{fig:rdfs_one} B). These two subpeaks are naturally summed into the one main CIP peak in the cation-oxygen RDF (Figure \ref{fig:rdfs_one} F, G). In case of potassium both these states are also realized, but they just are not resolved in cation-carbon RDF, however distinctive skewness in CIP potassium peak in Figure \ref{fig:rdfs_one}F is visible.

\begin{table*}
  \caption{Mean numbers of ions occupying CIP and SIP contact states with carboxylate}
  \label{tbl:all}
  \begin{tabular*}{\linewidth}{l@{\extracolsep{\fill}}cccccc}
  \hline
Parameters	&	ion	&	$r_{CIP}$ (nm)\textsuperscript{\emph{a}}	&	$r_{SIP}$ (nm)\textsuperscript{\emph{a}}	&	$N_{CIP}$\textsuperscript{\emph{b}}
&	$N_{SIP}$\textsuperscript{\emph{b}}	&	$N_{CIP+SIP}$\textsuperscript{\emph{b}}	\\
    \hline                      
KB FF	&	K$^{+}$	&	0.44	&	0.69	&	 0.21 $\pm$  0.00	&	 0.83 $\pm$  0.01	&	 1.04 $\pm$  0.01	\\
KB FF	&	Na$^{+}$	&	0.38	&	0.62	&	 0.12 $\pm$  0.01	&	 0.79 $\pm$  0.02	&	 0.91 $\pm$  0.03	\\[2pt]
Aqvist FF	&	K$^{+}$	&	0.41	&	0.66	&	 0.12 $\pm$  0.01	&	 0.74 $\pm$  0.01	&	 0.86 $\pm$  0.02	\\
Aqvist FF	&	Na$^{+}$	&	0.38	&	0.63	&	 0.21 $\pm$  0.01	&	 0.68 $\pm$  0.02	&	 0.89 $\pm$  0.03	\\
    \hline
  \end{tabular*}
  
  \textsuperscript{\emph{a}} Positions of the minima in the RDFs (Figure \ref{fig:rdfs_one} D,E) that corresponds to the upper bound of the CIP or SIP state;
  \textsuperscript{\emph{b}} Calculated with eq. (\ref{eq:rdf}) from RDFs shown in Figure \ref{fig:rdfs_one} D,E (1M salt system), for $N_{CIP}$ integration in (\ref{eq:rdf}) is from $0$ to $r_{CIP}$, for $N_{SIP}$ -- from $r_{CIP}$ to $r_{SIP}$, for $N_{CIP+SIP}$ -- from $0$ to $r_{SIP}$. Statistical errors obtained with block averaging (5 blocks).
\end{table*}

The first peak for sodium is higher than corresponding peak for potassium in all RDFs for both force fields. However, the height of the peak in RDF does not unambiguously define a strength of binding, because more significant value is the mean occupancy of bound state. The mean number of of ions bound in a particular coordination state (i.e. CIP or SIP) may be obtained by integration of the corresponding RDF peak as follows
\begin{equation}
\label{eq:rdf}
N_{bound} = \int_{r_1}^{r_2} dr\, 4\pi r^2 g(r),
\end{equation}
where $g(r)$ -- is a radial distribution function, $r_1$ and $r_2$ defines a boundaries of the bound state. We made such calculations with RDF between ion and carboxyl carbon atom, because integration of the ion-oxygen RDF to oxygen is complicated be presence of two oxygen atoms in carboxylat. The positions of the minima following the first and second main peaks on the RDFs (Figure \ref{fig:rdfs_one} D,E) were chosen as the upper bounds of the CIP and SIP bound states, respectively. These boundary distances ($r_{CIP}$ and $r_{SIP}$) and the corresponding mean numbers of ions for CIP and SIP states ($N_{CIP}$ and $N_{SIP}$) are summarized in Table \ref{tbl:all}. From the comparison of the $N_{CIP}$ values, it appears that for KB parameters in CIP configuration potassium surprisingly binds stronger than sodium, despite the fact that first peak on RDF for sodium is higher (see Figure \ref{fig:rdfs_one} D). For Aqvist FF situation is the opposite to KB FF, and CIP state is about 2 times more occupied in case of sodium than for potassium, which is in qualitative agreement with experimental data that sodium-carboxylate binding is stronger \cite{daniele2008weak,aziz2008cation}. We note that there is no contradiction in such deviations from experiment for KB FF, because this FF was parametrizes to reproduce the thermodynamic properties of acetate salt solutions using the Kirkwood-Buff theory, in which the integral of the RDF over the entire range is of central importance, while for $N_{CIP}$ integration is limited by the interval $[r_1,r_2]$ \cite{hess2009cation}. Mean number of cations in SIP are comparable, and this state is slightly more occupied in case of K$^+$ ions for both FFs. Interestingly, the values of CIP state occupancies for Na$^+$ and K$^+$ are swapped for two force fields (cf. $N_{CIP}$ in Table \ref{tbl:all}), it appears that the CIP binding strength of potassium in KB FF and sodium in Aqvist FF coincide and vice versa. Thus it is a clear starting point to check whether the strength of ion binding, and consequently the stability of cation-carboxylate CIP, affects the interactions of positively charged protein groups with carboxylate. 

\begin{figure*}[ht!]
  \includegraphics[width=1.0\linewidth]{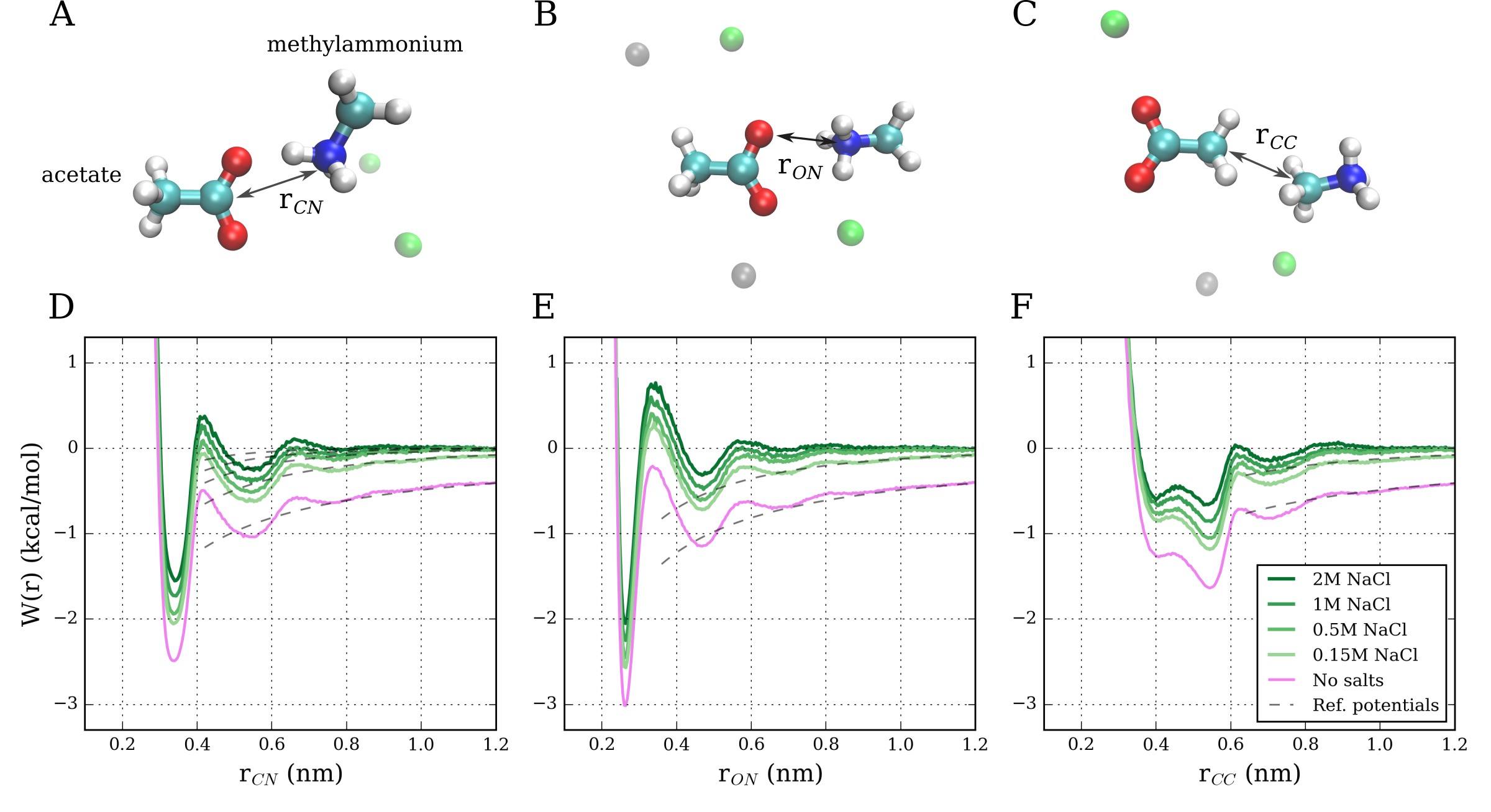}
  \caption{Potentials of mean forces (PMFs) between acetate and methylammonium. (A-C) Snapshots from a simulation of the acetate and methylammonium in 1M NaCl solution, showing different reaction coordinates. (A) Coordinate between the carbon atom of carboxyl group of acetate and nitrogen atom of amino-group of methylammonium ($r_{CN}$), PMF along this coordinate is shown in panel D. (B) Coordinate between the nitrogen atom of methylammonium and the nearest oxygen atom of acetate ($r_{ON}$), corresponding PMF is shown in panel E. (B) Coordinate between the carbon atoms of methyl groups of actetate and methylammonium ($r_{CC}$), corresponding PMF is shown in panel F. (D-F) PMFs along different coordinates, obtained from MD trajectories computed with KB FF ion parameters. PMF for salt free system is shown with violet line, for systems with salts -- darker green PMF curves correspond to increase in NaCl concentration. Reference continuum electrostatic potentials are shown with dashed lines, not all shown for clarity. Nitrogen atom of methylammonim is shown in blue, other atoms are colored as in Figure \ref{fig:rdfs_one}, green and gray spheres in the background -- are sodium and chlorine dissolved ions.}
    \label{fig:pmfs_conc}
\end{figure*}

\subsection{Effect of ions on charged group association}
To tests the ion-specific effects on the interactions between model protein charge groups we calculated potentials of mean force (PMFs) between the mentioned above acetate molecule and methylammonium molecule in presence of different amounts of NaCl or KCl for both force fields. While acetate contains an anionic carboxyl group $COO^-$, present in acidic amino acid residues, methylammonium contains a cationic amino group $NH_3^+$, which can be found in the LYS residue and at the N-termini of the proteins \cite{thomas2006direct}. PMFs were calculated along different methylammonium-acetate intermolecular distances from free MD trajectories, as detailed in Methods section. Considered reaction coordinates are depicted in Figure \ref{fig:pmfs_conc} A, B and C, and corresponding PMFs, are shown in Figure \ref{fig:pmfs_conc} D, E and F (for systems with different NaCl salt concentrations and KB FF). First minimum in PMF along carbon-nitrogen coordinate (Figure \ref{fig:pmfs_conc} A and D, $r_{CN}\sim0.34$ nm) and in PMF along oxygen-nitrogen coordinate (Figure \ref{fig:pmfs_conc} B and E, $r_{ON}\sim0.26$ nm) constitutes the same direct charge-charge contact between amino- and carboxyl groups, which is of primary interest for the present study. A different picture is observed for PMF along the distance between carbon atoms of methyl groups (Figure \ref{fig:pmfs_conc} C and F). In this case first local minimum ($r_{CC}\sim0.4$ nm) corresponds to a direct hydrophobic contact of methyl groups. While the second deeper minimum ($r_{CC}\sim0.55$ nm) is attributed to conformations for which there is no direct hydrophobic contact, but the charge-charge contact is established, as it was shown previously by Thomas and Elcock \cite{thomas2006direct}.

For all PMFs, it is clearly seen how the increase in salt concentrations gradually screens electrostatic attraction between methylammonium and acetate. Furthermore, starting from sufficient separations obtained PMF curves accurately follow analytical potentials of continuum electrostatics (eq. (\ref{eq:scr_coulomb})) depicted with dashed lines in the figure. Most clearly it is seen for carbon-nitrogen and carbon-oxygen coordinates, which reflect the distance between charged groups of the molecules (Figure \ref{fig:pmfs_conc} D and E), especially for Coulomb potential for salt free system and screened Coulomb in case of 0.15M NaCl. It was found that PMFs already follow electrostatic potentials starting from $\sim$0.8 nm separations, while each PMF curve was fitted to corresponding reference potential at distances larger that 1.0 nm (fitting intervals were $[1.2, 1.5]$ for carbon-nitrogen coordinate and $[1.1, 1.4]$ in other cases). Interestingly, even for methyl-methyl reaction coordinate ($r_{CC}$) PMF curves reaches electrostatic potentials for large distances (Figure \ref{fig:pmfs_conc} F). These observations indicates the correctness of our choice of the reference potentials, even though the electrostatic interactions in MD are actually computed in the periodic system.

\begin{figure*}[ht]
  \includegraphics[width=0.8\linewidth]{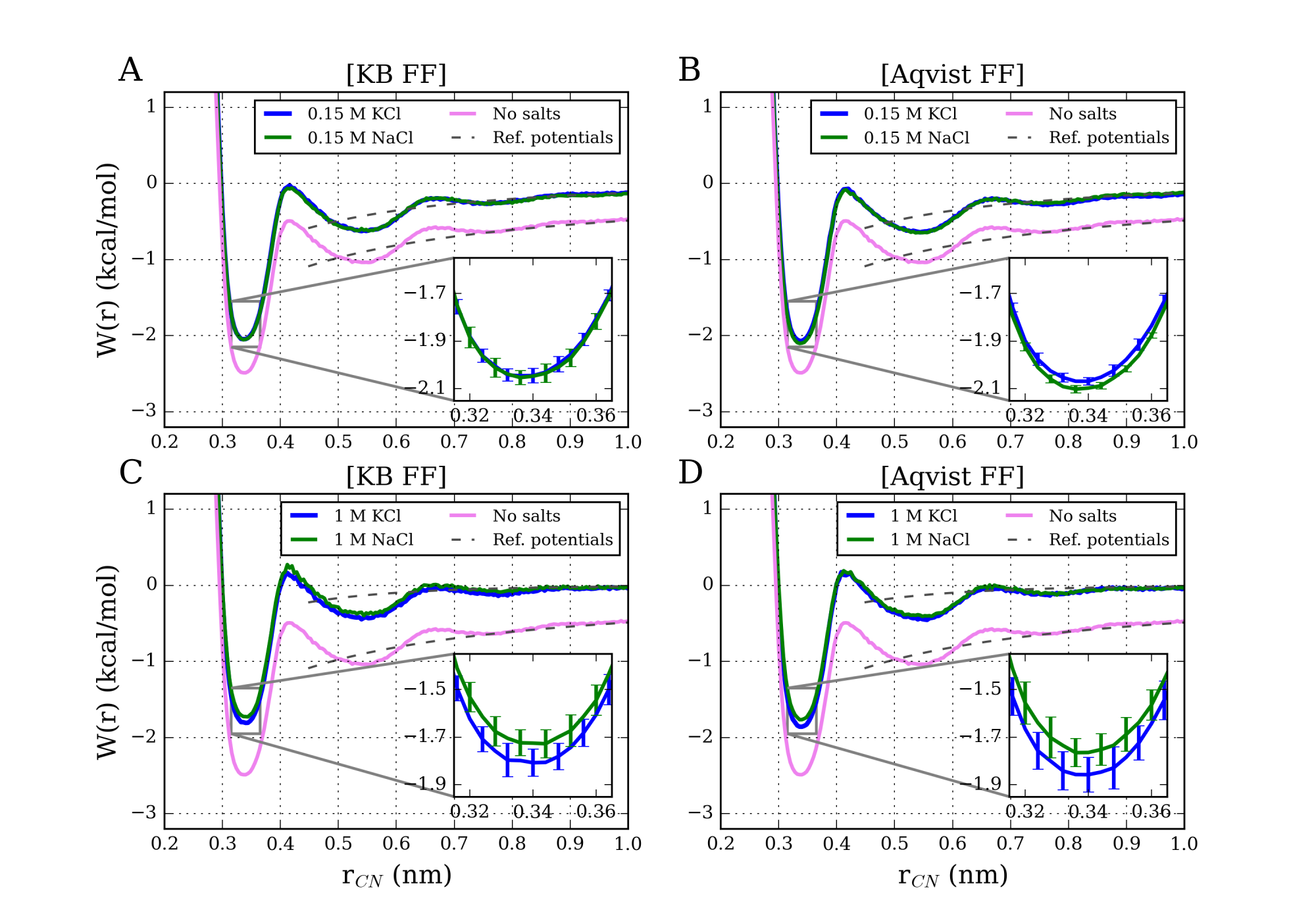}
  \caption{Comparison of the potentials of mean force (PMFs) between charged groups of acetate and methylammonium (along carbon-nitrogen coordinate), calculated in presence of either Na$^+$ or K$^+$ ions. (A,B) PMFs for systems with 0.15M concentration of NaCl or KCl salt. (C,D) PMFs for systems with 1M of NaCl or KCl. PMFs in panels A and C were obtained from MD simulations with KB FF ion parameters, in panels B and D -- with Aqvist FF. For systems containing Na$^+$ PMF curves are shown with green, for systems with K$^+$ -- blue, for salt free system -- violet. Reference continuum electrostatic potentials are shown with dashed lines.}
  \label{fig:pmfs_diff}
\end{figure*}

Comparison of carbon-nitrogen PMFs for systems which contain the same concentrations of either Na$^+$ or K$^+$ ions surprisingly did no reveal any pronounced ion-specific effects for both FFs -- PMF curves in case of potassium and sodium almost coincide (Figure \ref{fig:pmfs_diff}). From a closer look on the contact minimum in the insets on the Figure \ref{fig:pmfs_diff} (top panels -- 0.15M salt concentration, bottom panels -- 1M) it is seen that the difference in the minimum depending on the cation type is very small compared to kT (in present conditions kT$\approx0.6$ kcal/mol) and does not exceed the standard deviations, obtained with block averaging. The similar picture is observed for all studied salt concentrations (Figure \ref{fig:pmfs_diff} and Figures S1-S5 in SI). It should be noted that in the particular important case of physiological 0.15 M salt concentration ion-specific effects appeared to be negligible regardless of the FF variant used (Figure \ref{fig:pmfs_diff} A, B). Although for 1M slightly stronger charge-charge contact is observed in presence of K$^+$ ions, the difference nevertheless does not exceed the error level (Figure \ref{fig:pmfs_diff} C, D). The results are overall the same for both ion FFs, and thus we may conclude that the difference in the strength of ion-carboxylate CIP binding (see Table \ref{tbl:all}) does not lead to a significant ion-specific influence on the charge-charge contact between carboxyl and amino-group, at least in case of methylammonium and acetate molecules.

\begin{figure}[ht]
  \includegraphics[width=0.9\linewidth]{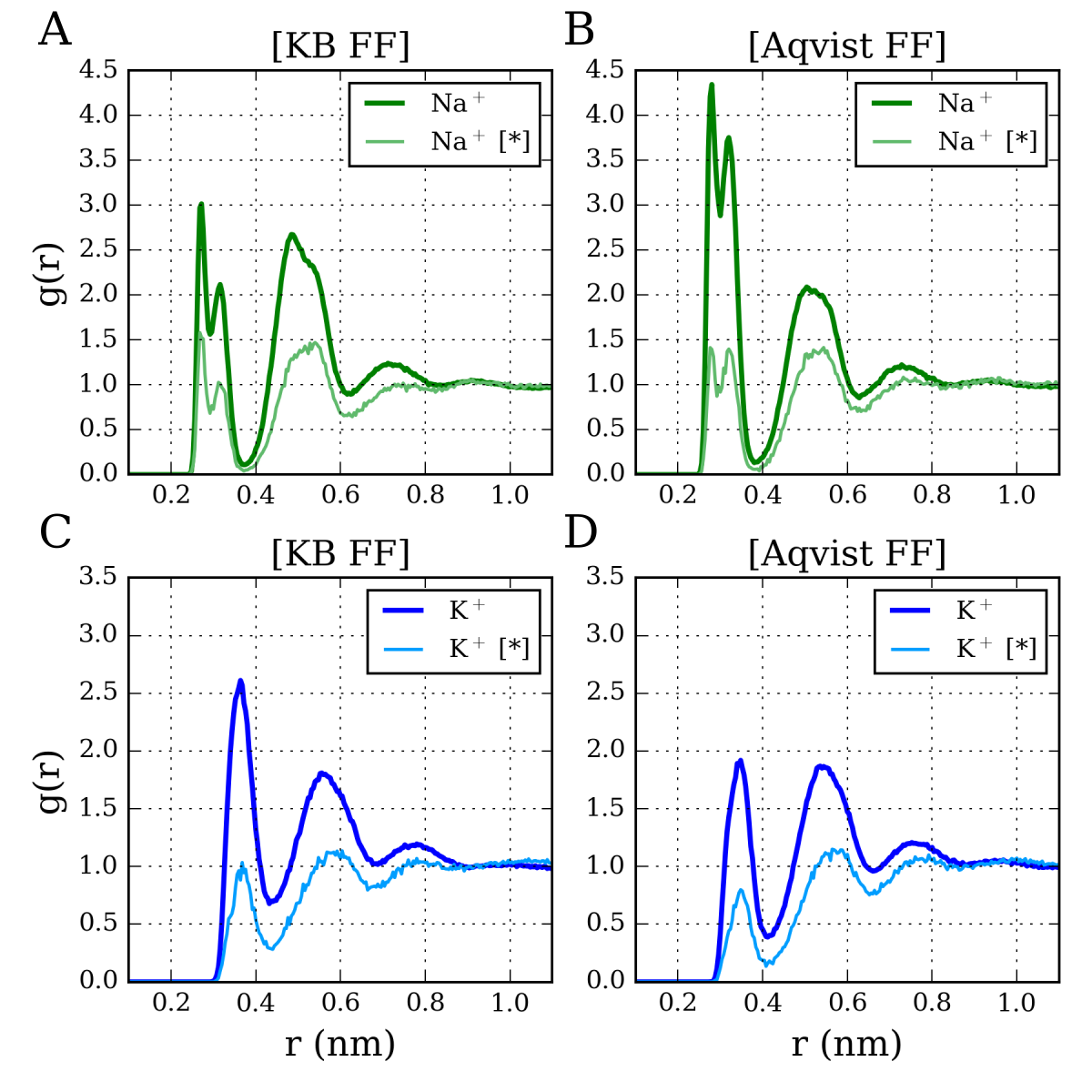}
  \caption{Influence of the charge-charge contact between carboxylate and amino-group on the cation coordination around  carboxylate. Radial distribution functions (RDFs) between carboxyl carbon atom of acetate and cations for Na$^+$ are shown with green lines (A,B), for K$^+$ -- with blue lines (C,D). In each panel lighter lines marked with asterisk [*] show RDFs computed from configurations in which the methylammonium-acetate charge-charge contact is established (i.e. $r_{CN}\leq0.42$ nm) in simulations of 1M salt systems. Darker lines show RDFs for 1M salt systems without methylammonium. RDFs in panels A and C were obtained from MD simulations with KB FF ion parameters, in panels B and D -- with Aqvist FF.} 
  \label{fig:rdfs_bounded}
\end{figure}

Interestingly, formation of potassium and sodium CIPs with carboxylate is still possible when this group participates in the charge-charge contact with amino-group. The bound state with a charge-charge contact corresponds to those conformations when the carbon-nitrogen distance is within the first minimum in the carbon-nitrogen PMF, which has the upper distance boundary at the desolvation maximum $r_c=0.42$ nm (see Figure \ref{fig:pmfs_diff}).  RDFs marked with an asterisk ($[*]$) and shown with lighter lines in Figure \ref{fig:rdfs_bounded} were calculated from a subset of configurations with established charge-charge contact between methylammonium and acetate (i.e. when $r_{CN}\leq0.42$ nm). Compared to coordination with free carboxylate (darker lines in Figure \ref{fig:rdfs_bounded}) when there is a contact with an amino-group the occupancies of the CIP and SIP state are inhibited, but does not go to zero. The CIP occupancies decrease approximately by a factor of 2-4 for the studied salt concentrations, which corresponds to a difference in the ion-carboxylate binding free energy of only about kT (see Figure S6-S8 in SI). This indicates that carboxylate should not be considered as a specific ligand-binding site that can be associated with either an ion or a protein cationic group. This reasoning partly explains why the difference in ion-cabroxylate binding strength together with distinct local spatial distributions of Na$^+$ or K$^+$ ions (see Figure \ref{fig:rdfs_bounded}) has almost no effect on charge-charge contact between methylammonium and acetate.

\subsection{Ionic strength dependence}
The above comparison of charge-charge contact PMF minima does not reveal any significant dependence on the ion type for the same salt concentrations. Moreover, the change in salt concentration only shifts this minimum, while its shape remains unaffected by the surrounding ions -- PMFs minima for different salt concentrations completely overlap when superimposed (see Figure S9 in SI). Thus we do not observe any local effects of dissolved ions on the model charged group interactions, and the main ion influence comes from non-local electrostatic screening. Such screening produced by ionic atmosphere in the solution is generally not ion-specific and is mostly determined by the ionic strength ($I$), which in case of NaCl and KCl solutions simply equal to the salt concentration.

\begin{figure}[ht]
  \includegraphics[width=0.9\linewidth]{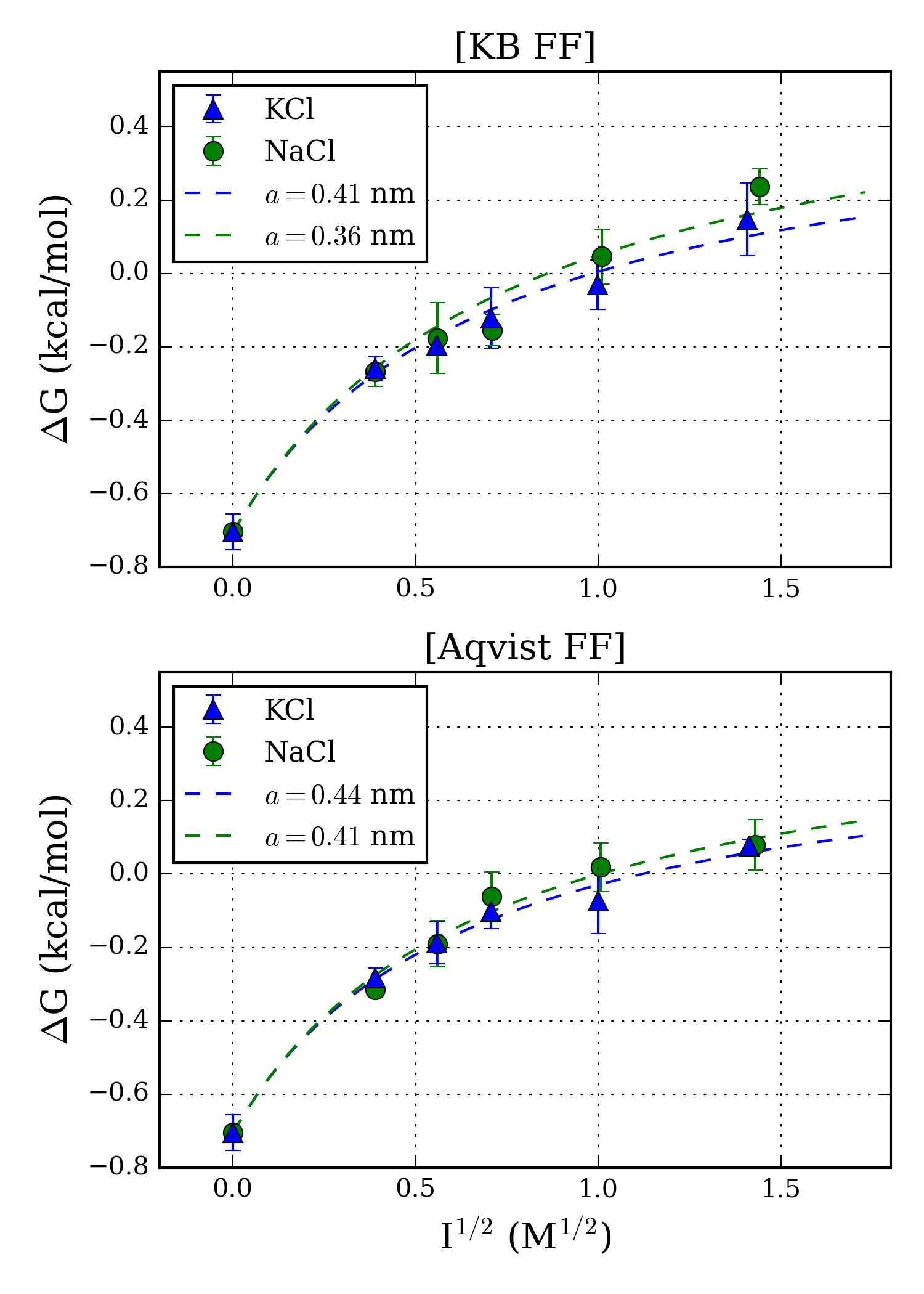}
  \caption{Free energies of charge-charge contact binding between methylammonium and acetate for different ionic strength of the solution $I$. Values of $\Delta G$ computed from MD simulations of NaCl solution systems are shown with green circles, for KCl solutions -- with blue triangles; errors obtained with block averaging. Top panel -- results for KB FF, bottom panel -- for Aqvist FF. Best fits of the proposed theory (\ref{eq:final_eq}) to the data for NaCl solutions are shown with green dashed lines, for KCl -- blue dashed lines. Fitted values of the parameter $a$ (distance of closest approach) are shown in legend for each FF.}
  \label{fig:dg}
\end{figure}

For quantitative comparison of the ion effects on the charge-charge contact, the binding free energies were calculated from carbon-nitrogen PMFs with eq. (\ref{eq:k_a}), with a bound state defined as described above when $r_{CN}$ is within a contact minimum (carbon-nitrogen distances are less than $r_c = 0.42$). The ionic strengths dependences of the charge-charge binding free energy are shown in Figure \ref{fig:dg}. As it was discussed in case of PMFs, there is also no pronounced ion-specific trend in $\Delta G$ values for both used FFs. Furthermore, values of $\Delta G$ at the same salt concentrations for different ion FFs coincide up to the statistical errors, except the case of 2M salt systems ($I^{1/2} \sim 1$ M$^{1/2}$). The difference for such a high salt concentration may appear due to the ion clustering artifacts that depend on the force field. 

The absence of local ionic effects means that the structure and energy of solute-solute and solute-solvent interactions in the bound state are not affected by surrounding ions, as indicated by the PMF minima overlap (Figure S9). In turn, non-local effects of ions may be described in such a way that the charged groups lose their favorable interactions with ionic cloud, when they form a charge-neutral bound state. In this considerations the apparent association constant and binding free energy ($K_a$ and $\Delta G$) in the system with certain salt concentration are related to the standard binding constant and free energy ($K_a^o$ and $\Delta G^o$) at infinite dilution (i.e. in system with no salts) with the following expression~\cite{marcus2006ion}
\begin{equation}
\label{eq:ka_actitity}
K_a = K_a^o\cdot \gamma_{\pm}^2/\gamma_{b}; \, \,  \Delta G = \Delta G^o - 2 RT\ln \gamma_{\pm} +RT \ln  \gamma_{b},
\end{equation}
where $\gamma_{\pm}$ and $\gamma_{b}$ are the mean activity coefficient of dissolved free species and activity coefficient of the bound complex, respectively; $R$ -- is a gas constant and $T$ -- temperature. Interactions with ionic surrounding results in the deviation of the activity coefficients from the ideal value equal to one, thus for charge-neutral complex it was set to unity ($\gamma_{b} =1$), which is a usual approximation \cite{marcus2006ion}. While for charged species interaction with ionic cloud is essential and the mean activity coefficient for cationic methylammonium and anionic acetate may be obtained within extended Debye-Huckel theory \cite{debye1954collected,marcus2006ion} as follows 
\begin{equation}
\label{eq:EDH}
\ln \gamma_{\pm} = - q \cdot  \frac{\kappa }{ 1 + a\cdot \kappa}; \, \, q = \frac{q_e^2}{2\varepsilon_0 k T},
\end{equation}
where $\kappa$ is a Debye parameter (see Methods), which is propositional to the square root of the solution ionic strength, $a$ -- is an adjustable parameter with dimension of length, that roughly reflects a distance of closest approach between ions; $k$ -- Bolzmann constant, $q_e$ -- elementary charge and $\varepsilon_0$ -- vacuum electric permittivity. Taking (\ref{eq:ka_actitity}) and (\ref{eq:EDH}) together gives the expression for ionic strength dependence of $\Delta G$ under the above assumptions
\begin{equation}
\label{eq:final_eq}
\Delta G = \Delta G^o + 2 RT q \cdot \frac{\kappa^o \cdot I^{1/2}}{1 + a\cdot \kappa^o \cdot I^{1/2}},
\end{equation}
where $\kappa^o=\sqrt{2000 N_a q_e^2 /(\varepsilon \varepsilon_0 k T)}$ -- is a coefficient which remains after explicitly taking the ionic strength $I^{1/2}$ dependence and $q$ -- parameter defined in (\ref{eq:EDH}). The value of $\Delta G^o$ is obtained from the PMF in the salt free system, and than in the equation (\ref{eq:final_eq}) remains only one unknown parameter -- $a$. Values of $a$ were separately obtained by fitting the expression (\ref{eq:final_eq}) to the data on the ionic strength dependence of $\Delta G$ for either Na$^+$ or K$^+$ in case of each FF (Figure \ref{fig:dg}). The resulted fitted curves (dashed lines in Figure \ref{fig:dg}) closely follow  $\Delta G$ values, calculated from the MD simulations. It appears that, the proposed simple theoretical description is consistent with the observed effect of dissolved ions on charge-charge binding free energy. This confirms our statement on the predominantly non-local character of ion influence, which is only taken into account by the theory.

Obtained values of parameter $a$ (see legend in Figure  \ref{fig:dg}), are larger than actual distances of closest approach of Na$^+$ or K$^+$ ions to the carboxylate and Cl$^-$ ions to the amino group  (see RDFs in Figure \ref{fig:rdfs_bounded} and Figure S10 in SI). However, in the theory $a$ rather has the meaning of an adjustable parameter
and only approximately reflects the actual minimum possible distance, especially in the present case since the charged groups are not spherically symmetric \cite{marcus2006ion}. Intriguingly, the fitted values of $a$ almost exactly match the value of theoretical parameter $q$ (for present conditions $q=0.41$ nm, cf. $a$ in Figure \ref{fig:dg}), which plays the role of the threshold distance for ion pairing in the Bjerrum approach \cite{marcus2006ion}. But additional research is required to assess the significance of this observation, which is beyond the scope of this work. 

We may nevertheless speculate that the obtained values of $a$ have some physical meaning, since they also closely follow the corresponding values of ion-carboxylate CIP boundaries (see $r_{CIP}$ in Table \ref{tbl:all}). The values of $a$ as well as $r_{CIP}$ are smaller in case of sodium for both FFs, which correlates with the fact that atomic radii for sodium is smaller than for potassium. In this regard the ability of Na$^+$ ion to closer approach the anionic group than K$^+$ ion leads to a slightly more favorable interactions with ionic cloud for the same ion concentration. This is reflected in the slight difference in the charged group activity, which however have no effect in physiological conditions, while may produce some minor ion-specific effect at high concentrations (see Figure \ref{fig:dg}), which can be important in certain cases \cite{dubina2013potassium}.

\section*{Conclusion}
The main physical reason of the different roles of the very similar potassium and sodium ions in the living systems still remains unclear. One of the proposed hypotheses relates ion effects to the specific ion-pairing with carboxylates which may interfere with the formation of the functional contacts (salt-bridges) between ubiquitous anionic carboxylic groups and cationic groups of proteins \cite{vrbka2006quantification,heyda2009ion,friedman2011ions}. This hypothesis is supported by experimental observations of a stronger binding between carboxyl group and sodium ion compared to potassium\cite{daniele2008weak,aziz2008cation,kherb2012role}. Molecular details of ion-specific coordination around the carboxylate have also been shown previously for Na$+$ and K$^+$ in a number of simulation studies, but a detailed picture of how these ions may affect protein group interactions is still lacking. In this work we systematically address this issue with MD simulations utilizing two force fields for ions. Surprisingly we did not observe any pronounced ion-specific effects of Na$+$ and K$^+$ ions on the iterations between model solutes containing charged groups of proteins. PMFs between anionic carboxyl group of actetate and cation amino-group of methylammonium molecule coincided within the statistical error for systems that contained same amounts of either NaCl or KCl. This result was not sensitive to the choice of ion FF parameters, although for KB FF it turned out that Na$^+$ ions bind to the carboxylate weaker than K$+$ ions, while for Aqvist FF situation is the opposite. Thus we may conclude that the strength of ion binding to the carboxylate do not determine the ion-specific effect on protein group interactions. Comparable results were demonstrated in resent work on effect of Li$+$ and Na$^+$ ions on the salt bridge formation between carboxyl and guanidinium groups \cite{pylaeva2018salt}. In this work we have further demonstrated that even when the charge-charge contact between model groups is established there is a room for ion coordination, and ion-carboxylate binding is not restricted. From these observations it follows that the binding of potassium and sodium ions with a single carboxylic group should not be considered in terms of a ligand-binding  mechanism -- weak ion pairing do not block carboxylate for salt bridging with other groups. In proteins however the dense and precise positioning of carboxylates may result in the specific binding site (i.e. in selectivity filters of ion channels \cite{catterall2017chemical,roux2017ion}), whose specificity is determined to a greater extent by the structural properties but not by the strength of ion-carboxylate binding. 

Our analysis of the binding between model protein groups in presence of different salt concentrations have demonstrated that the main effect of dissolved ions is non-local and does not appear at the charge-charge contact, but is due to interactions of the charged species with ionic cloud. We found that this effect is very well described with extended Debye-Huckel theory and that it is only weakly ion-specific. In this regard, it should be noted that in the case when carboxylates are located on the surface of a protein or lipid membrane, the transient biding of ions to carboxyl groups can change the mean surface charge, which leads to a change in the potential of the electric double layer \cite{zhang2014free}. In this case, one may expect that local ion-specific interactions with surface carboxylates will cause non-local change in the electric potential that affect the interaction with the surface, however this question may be the subject of a separate study.

\section*{Acknowledgments}
This work was supported by the Russian Science Foundation (project 21-72-30018). S.K. contribution was supported by the the Institute for Basic Science in Korea, Young Scientist Fellowship (IBS-R024-Y3-2022)



\bibliography{ms}

\clearpage
\pagebreak

%
%
\setcounter{equation}{0}
\setcounter{section}{0}
\setcounter{figure}{0}
\setcounter{table}{0}
\setcounter{page}{1}
\renewcommand{\theequation}{S\arabic{equation}}
\renewcommand{\thefigure}{S\arabic{figure}}
\renewcommand{\bibnumfmt}[1]{[S#1]}
\renewcommand{\citenumfont}[1]{S#1}

\renewcommand\thesection{SI.\Roman{section}}
\renewcommand\thefigure{S\arabic{figure}}
\renewcommand\theequation{S\arabic{equation}}
\renewcommand\thepage{S\arabic{page}}

\thispagestyle{empty}

\begin{widetext} 

\begin{center}
\vspace{20cm}
{\huge{Supporting Information}}
\vspace{1cm}

{\large{  \bf  Specific and non-specific effects of sodium and potassium ions on the interactions between model charged groups of proteins}}
\vspace{1cm}

\large{Ivan N. Terterov$^{1}$, Sergei V. Koniakhin$^{2}$, Alexey A. Bogdanov$^{3}$ }
\vspace{1cm}

{$^{1}$ School of Physics and Engineering, ITMO University, St. Petersburg, Russia}

{$^{2}$ Center for Theoretical Physics of Complex Systems, Institute for Basic Science (IBS), Daejeon 34126, Republic of Korea}

{$^{3}$ Saint-Petersburg Clinical Scientific and Practical Center of Specialized Types of Medical Care (Oncological), St. Petersburg, Russia}
\vspace{1cm}
\end{center}

Comparison of the potentials of mean force (PMFs) between charged groups of acetate and methylammonium along carbon-nitrogen coordinate ($r_{CN}$), calculated in presence of various concentrations of either NaCl or KCl salt (Figures \ref{sfig:pmfs_01m}-\ref{sfig:pmfs_2m}). Superpositions of PMFs for various salt concentrations along the main minimum are shown in Figure \ref{sfig:pmfs_all_min}.

Mean numbers of ions coordinated in CIP (Figure \ref{sfig:a_m_n_cip_bounded}) or SIP (Figure \ref{sfig:a_m_n_sip_bounded}) states during MD simulations of systems with different salt concentrations. The ratios of the mean numbers of ions coordinated in CIP or SIP during whole MD simulation to the corresponding mean numbers for configureions with esrablished charge-charge contact are shown in Figure \ref{sfig:a_m_n_compare}.

Influence of the charge-charge contact between carboxylate and amino-group on the chlorine coordination around amino-group of methylammonium is shown in Fugure \ref{sfig:a_m_rdfs_bounded_CL_nz}.

\begin{figure}[ht!]
\centering
  \includegraphics[width=1.0\textwidth]{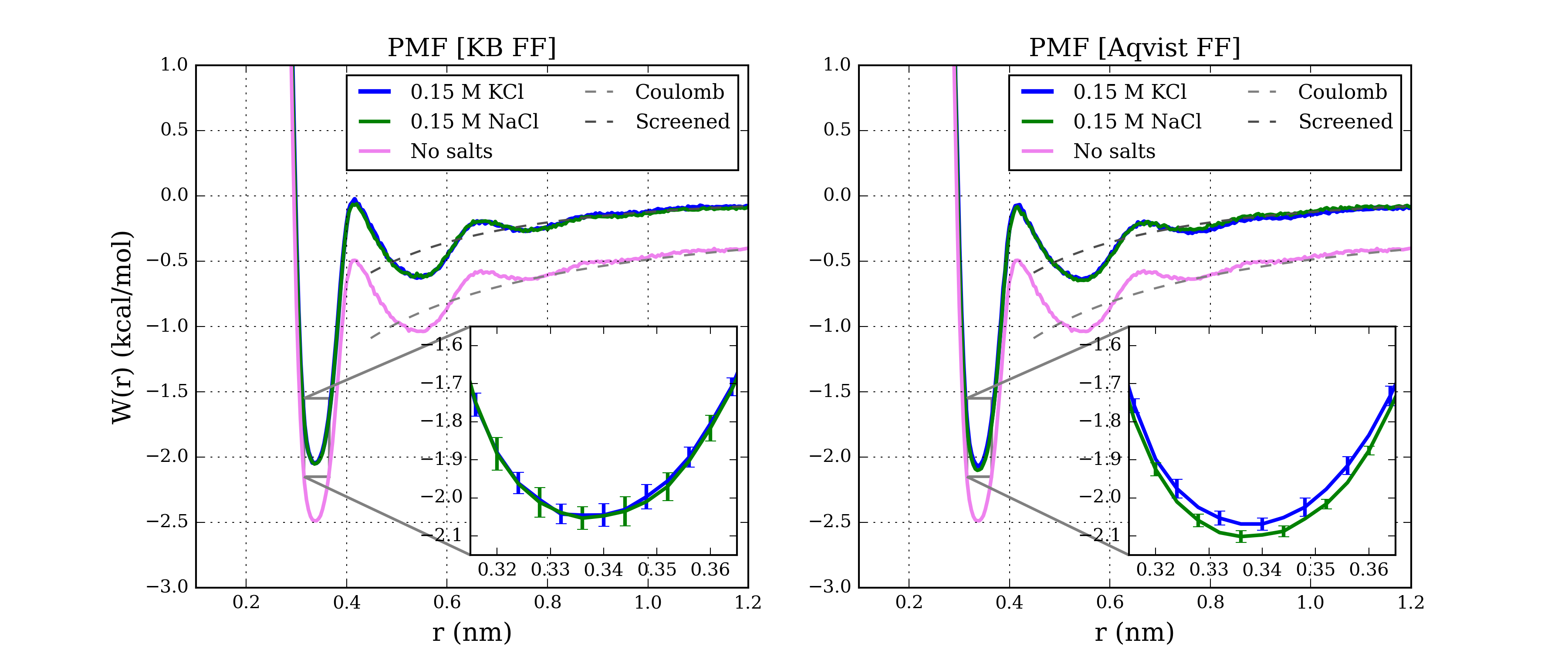}
  \caption{Comparison of the potentials of mean force (PMFs) between charged groups of acetate and methylammonium (along carbon-nitrogen coordinate), calculated in presence of 0.15M of either NaCl or KCl salt. PMFs in left panel were obtained from MD simulations with KB FF, in right panel -- with Aqvist FF. For systems containing Na$^+$ PMFs are shown with green lines, for systems with K$^+$ -- blue lines, for salt free system -- violet. Reference continuum electrostatic potentials are shown with dashed lines.}
  \label{sfig:pmfs_01m}
\end{figure}

\begin{figure}
\centering
  \includegraphics[width=1.0\textwidth]{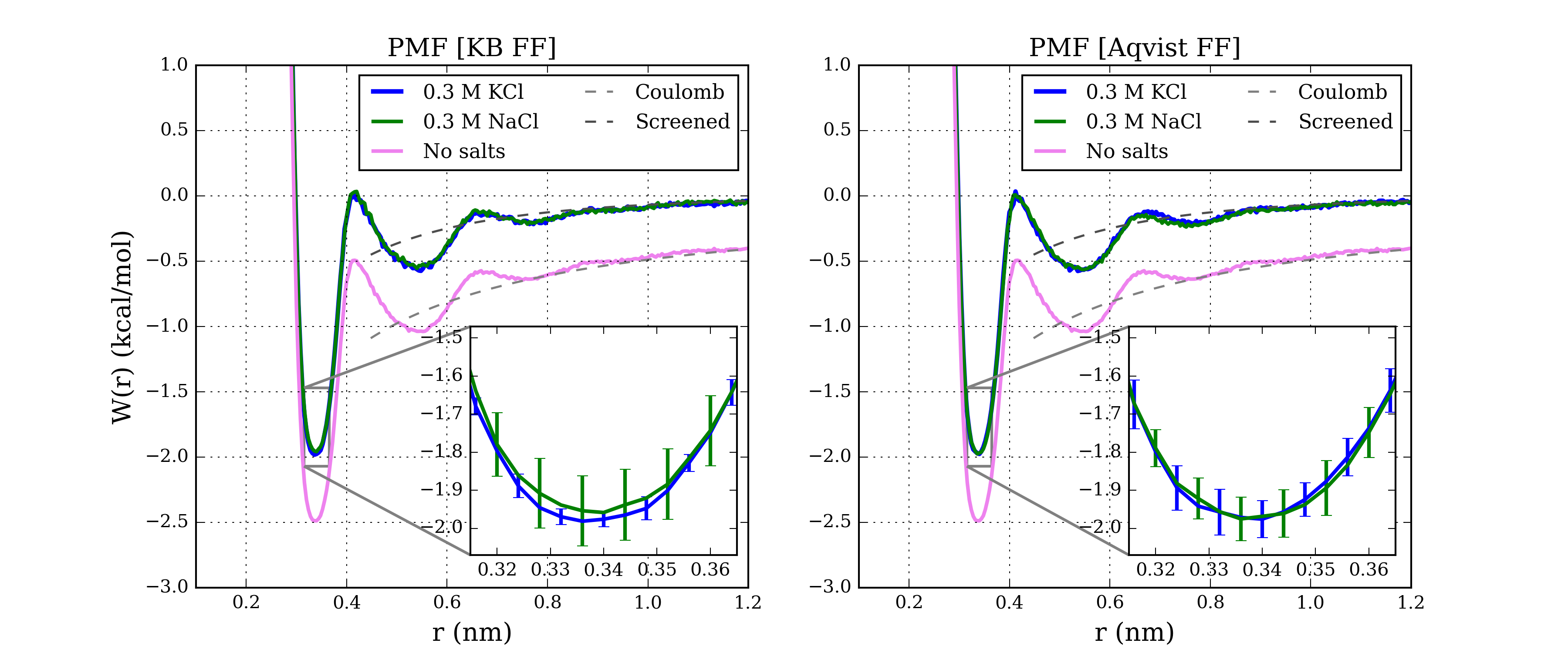}
  \caption{Comparison of the potentials of mean force (PMFs) between charged groups of acetate and methylammonium (along carbon-nitrogen coordinate), calculated in presence of 0.3M of either NaCl or KCl salt. All designations are the same as in the Figure \ref{sfig:pmfs_01m}}
  \label{sfig:pmfs_03m}
\end{figure}

\begin{figure}
\centering
  \includegraphics[width=1.0\textwidth]{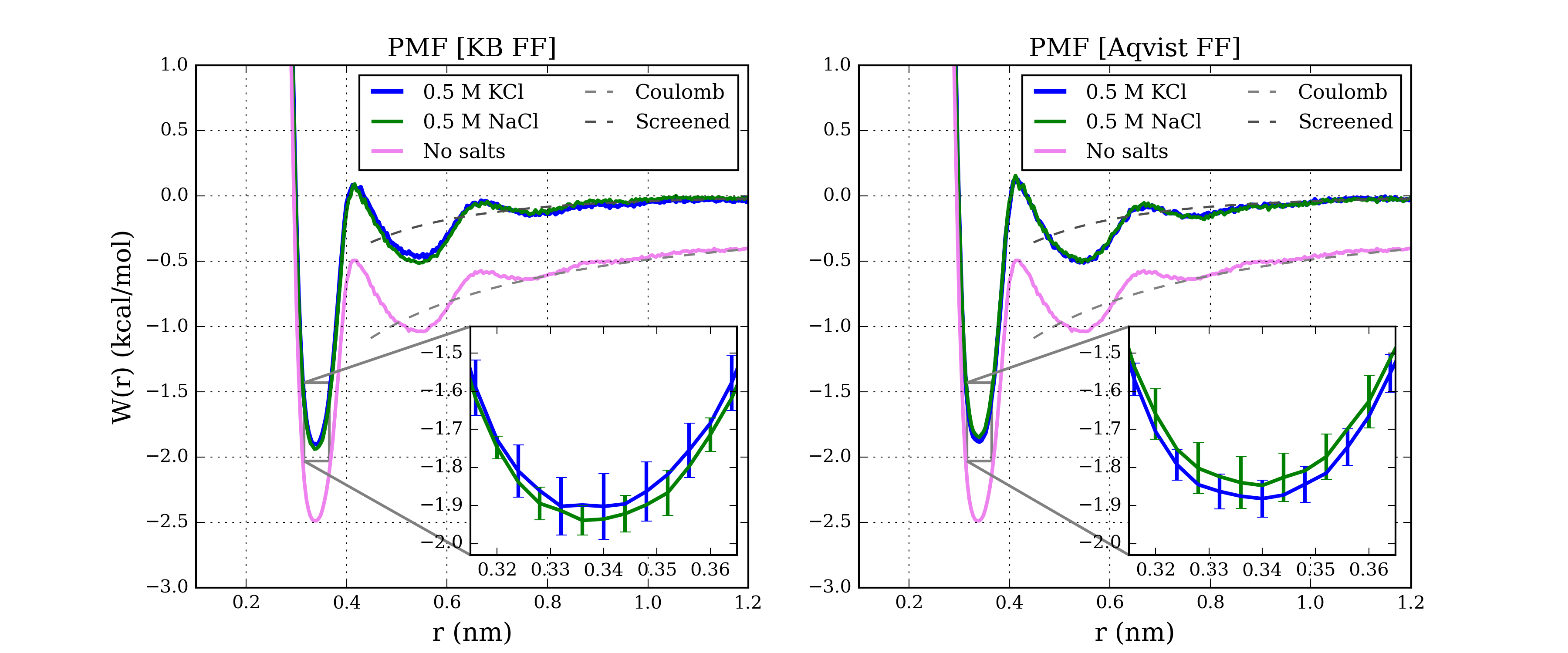}
  \caption{Comparison of the potentials of mean force (PMFs) between charged groups of acetate and methylammonium (along carbon-nitrogen coordinate), calculated in presence of 0.5M of either NaCl or KCl salt. All designations are the same as in the Figure \ref{sfig:pmfs_01m}}
  \label{sfig:pmfs_05m}
\end{figure}

\begin{figure}
\centering
  \includegraphics[width=1.0\textwidth]{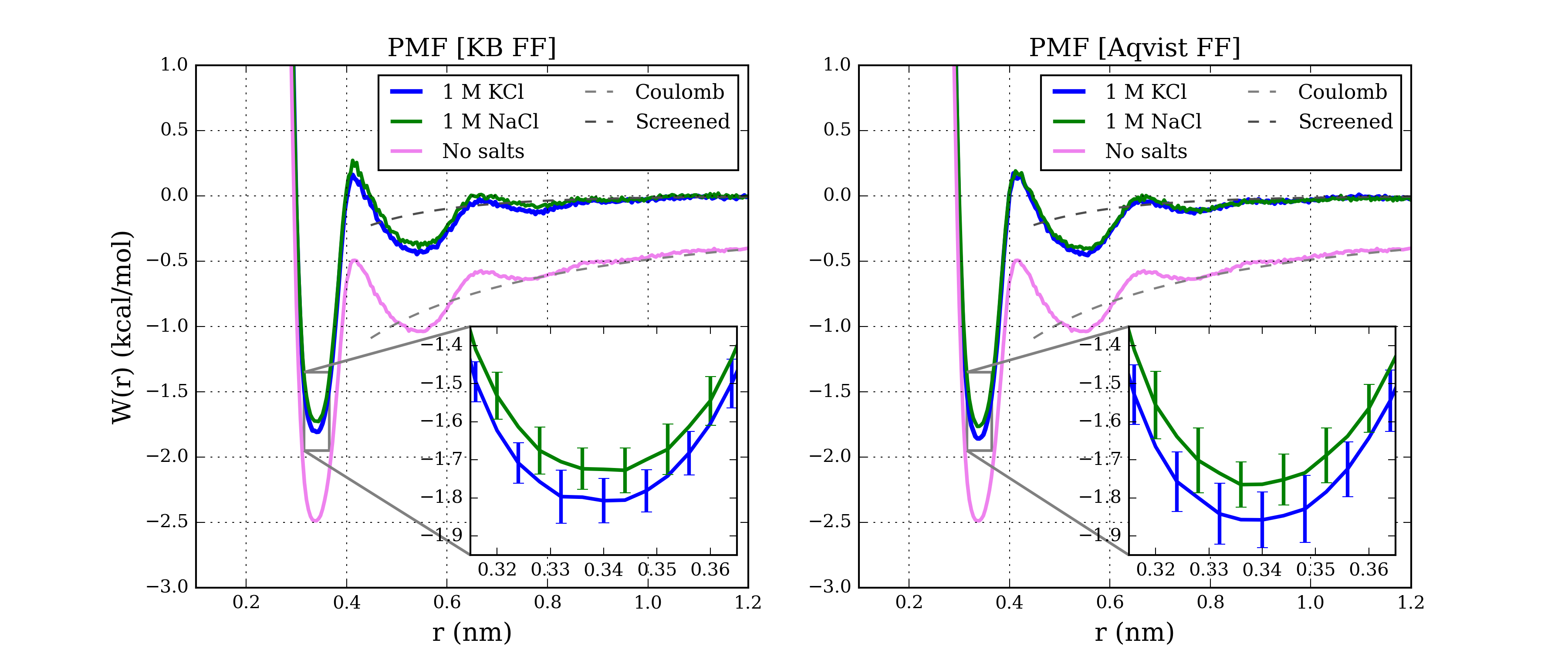}
  \caption{Comparison of the potentials of mean force (PMFs) between charged groups of acetate and methylammonium (along carbon-nitrogen coordinate), calculated in presence of 1M of either NaCl or KCl salt. All designations are the same as in the Figure \ref{sfig:pmfs_01m}}
  \label{sfig:pmfs_1m}
\end{figure}

\begin{figure}
\centering
  \includegraphics[width=1.0\textwidth]{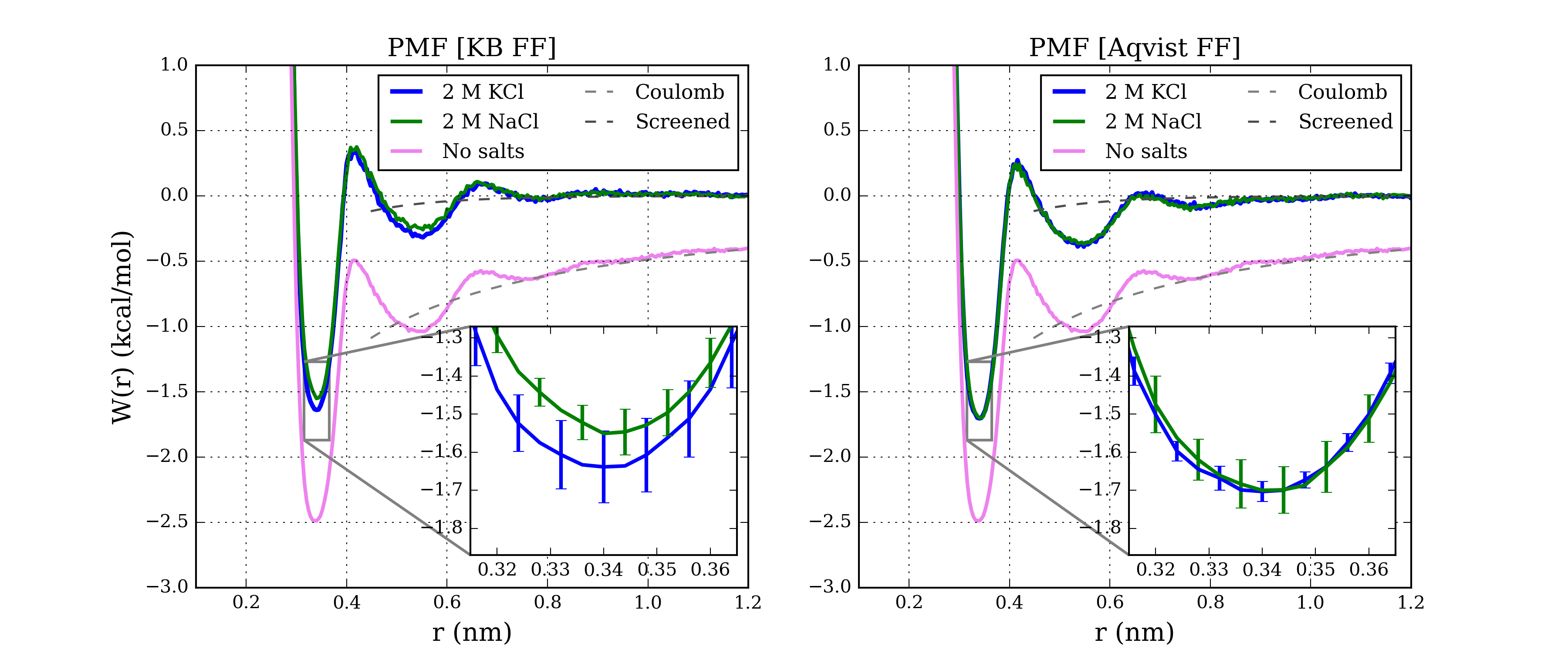}
  \caption{Comparison of the potentials of mean force (PMFs) between charged groups of acetate and methylammonium (along carbon-nitrogen coordinate), calculated in presence of 2M of either NaCl or KCl salt. All designations are the same as in the Figure \ref{sfig:pmfs_01m}}
  \label{sfig:pmfs_2m}
\end{figure}

\begin{figure}
\centering
  \includegraphics[width=0.6\textwidth]{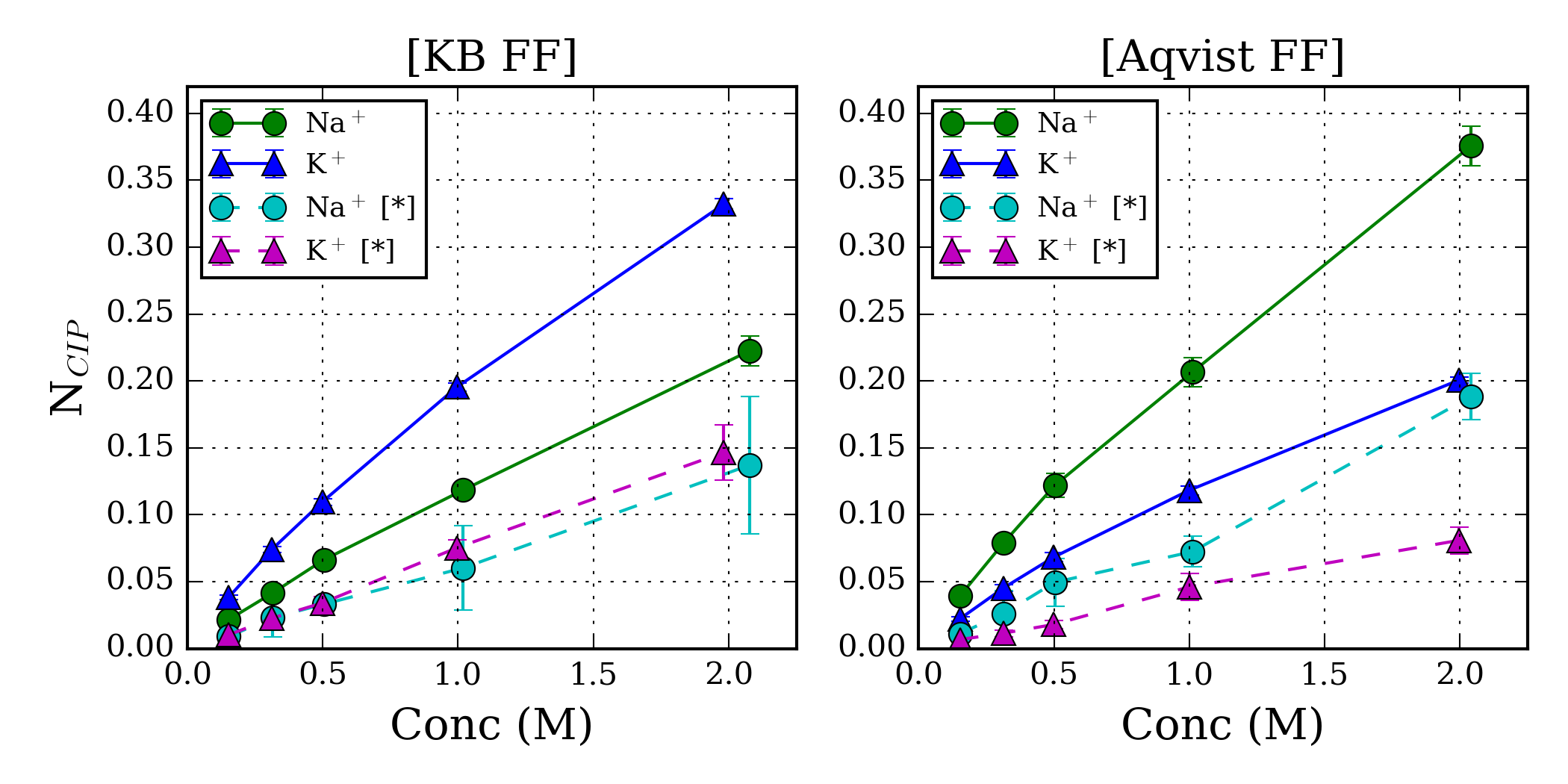}
  \caption{Mean number of ions coordinated in CIP state during MD simulations of systems with different salt concentrations. In each panel data marked with asterisk [*] shows $N_{CIP}$ values, computed from configurations in which the methylammonium-acetate charge-charge contact is established (i.e. $r_{CN}\leq0.42$ nm). Errors were obtained with block averaging.}
  \label{sfig:a_m_n_cip_bounded}
\end{figure}

\begin{figure}
\centering
  \includegraphics[width=0.6\textwidth]{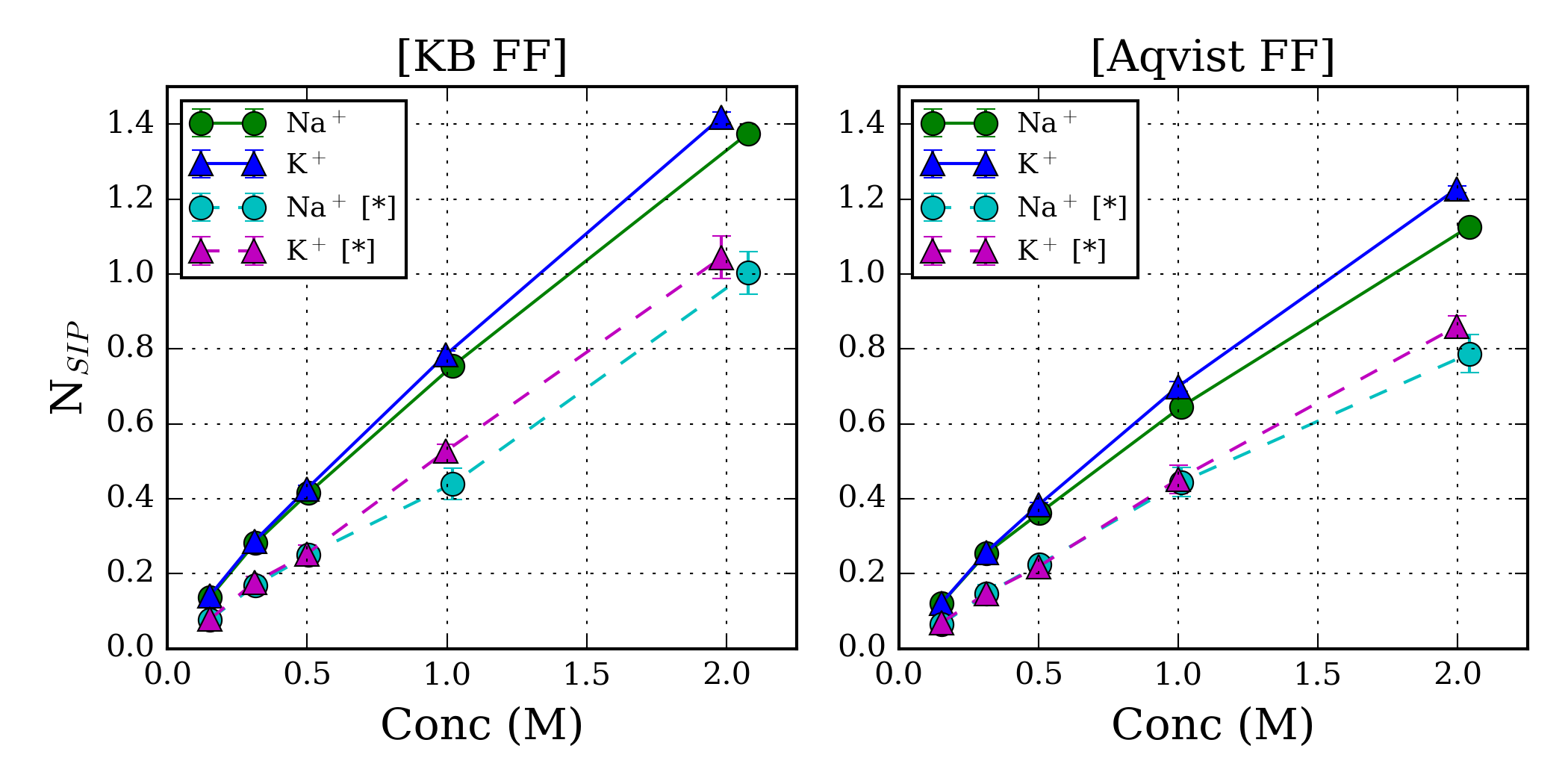}
  \caption{Mean number of ions coordinated in SIP state during MD simulations of systems with different salt concentrations. In each panel data marked with asterisk [*] shows $N_{SIP}$ values, computed from configurations in which the methylammonium-acetate charge-charge contact is established (i.e. $r_{CN}\leq0.42$ nm). Errors were obtained with block averaging.}
  \label{sfig:a_m_n_sip_bounded}
\end{figure}

\begin{figure}
\centering
  \includegraphics[width=0.5\textwidth]{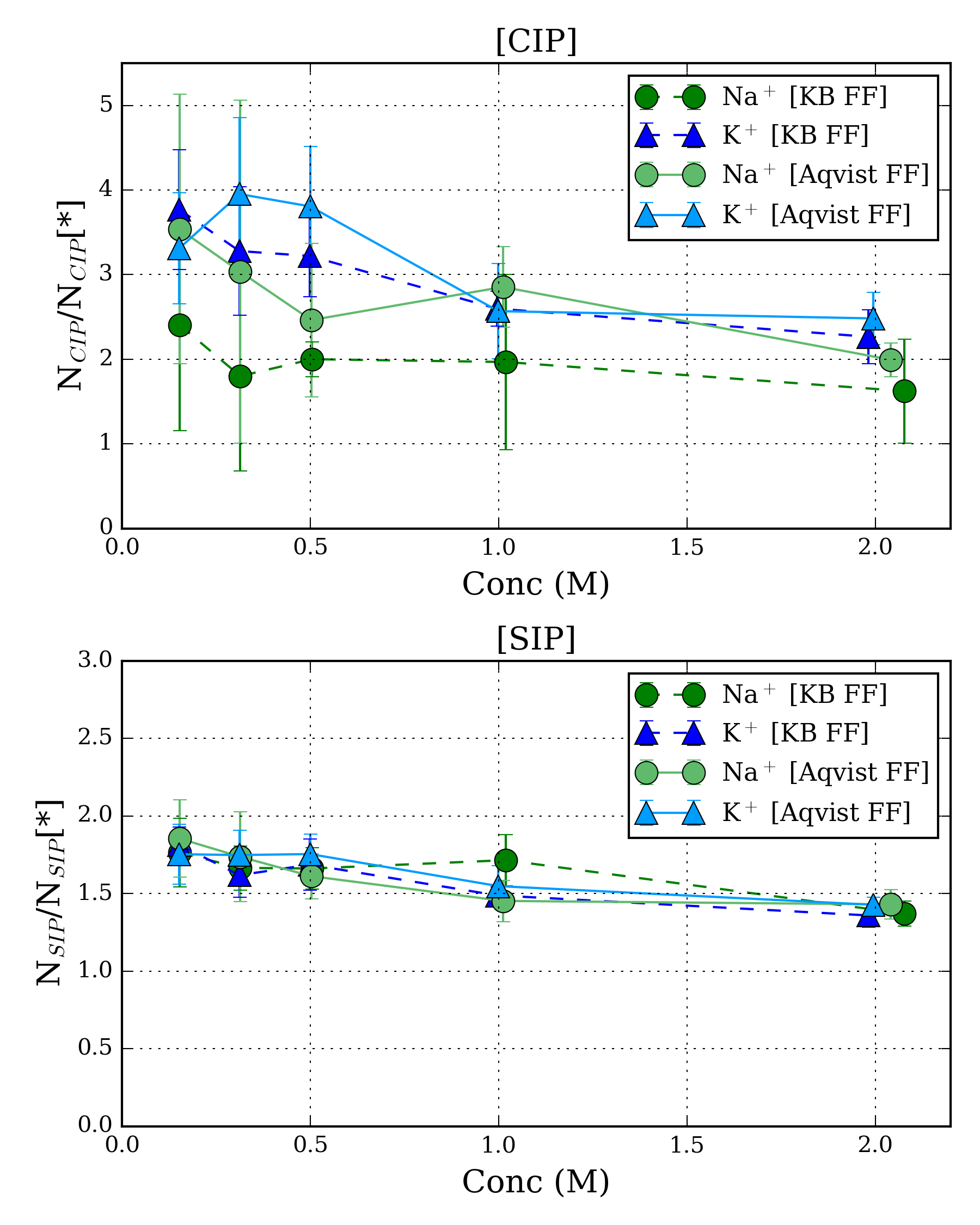}
  \caption{Ratios of the mean numbers of ions coordinated in CIP (top panel) or SIP (bottom panel) state during the entire MD simulations to the mean numbers of ions coordinated in  configurations in which the methylammonium-acetate charge-charge contact is established (i.e. $r_{CN}\leq0.42$ nm)}
  \label{sfig:a_m_n_compare}
\end{figure}

\begin{figure}
\centering
  \includegraphics[width=0.8\textwidth]{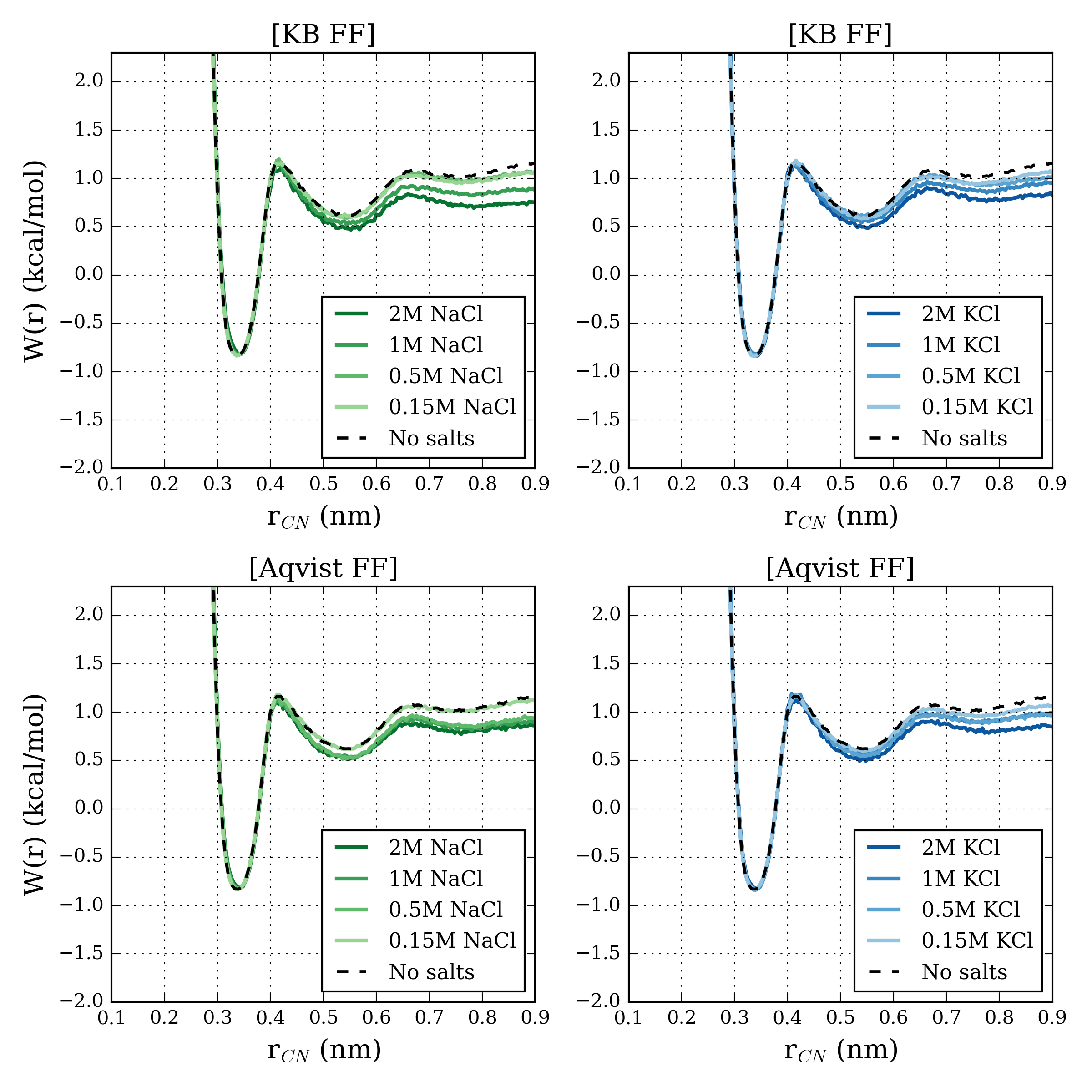}
  \caption{Comparison of potentials of mean forces between charged groups of acetate and methylammonium (along carbon-nitrogen coordinate), for different salt concentrations, when all curves are superimposed along the main minimum ($0.3\leq r_{CN}\leq0.42$ nm).}
  \label{sfig:pmfs_all_min}
\end{figure}

\begin{figure}
\centering
  \includegraphics[width=0.6\textwidth]{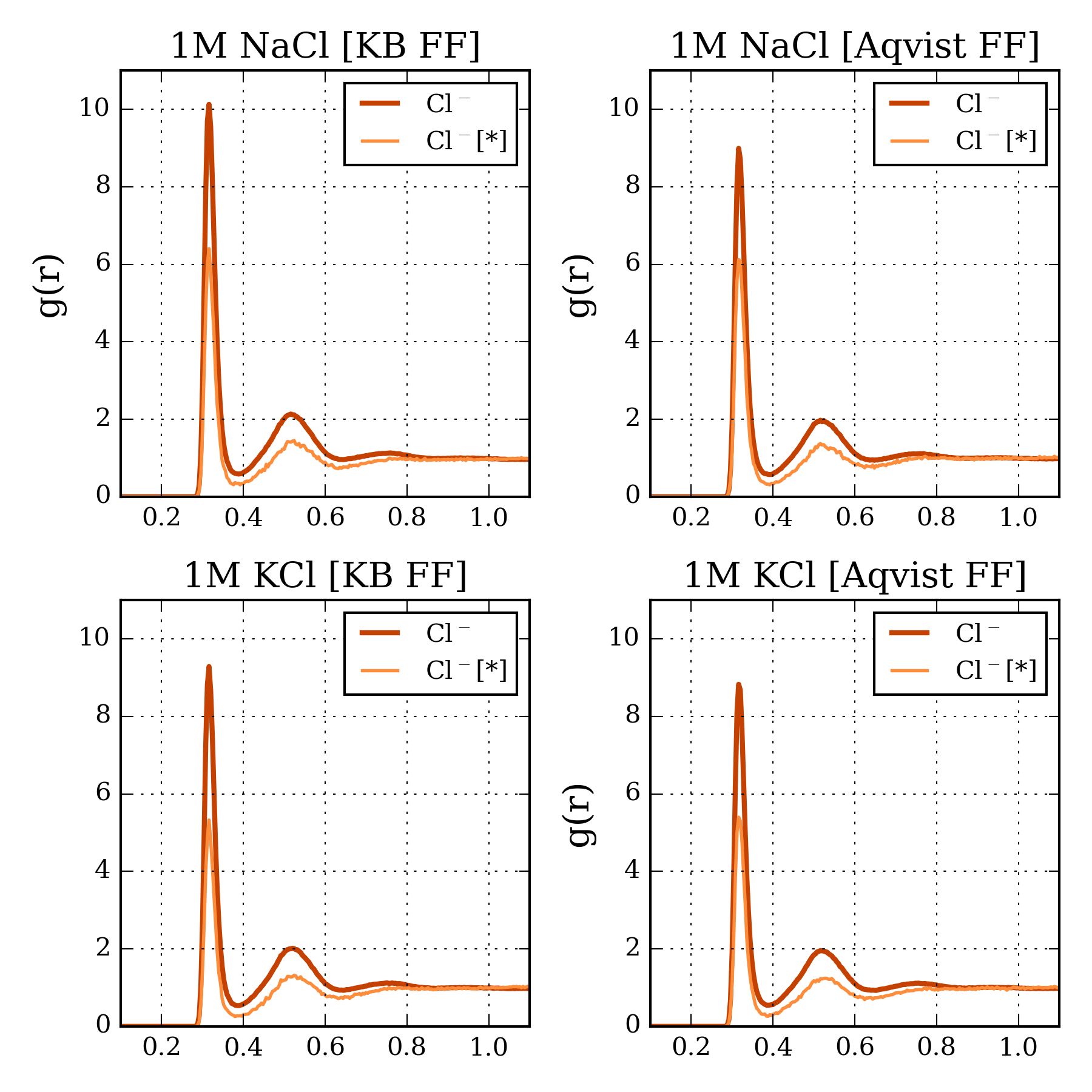}
  \caption{Influence of the charge-charge contact between carboxylate and amino-group on the chlorine coordination around  amino-group. Radial distribution functions (RDFs) between nitrogen atom of methylammonium and Cl$^-$ are shown with orange lines. In each panel lighter lines marked with asterisk [*] show RDFs computed from configurations in which the methylammonium-acetate charge-charge contact is established (i.e. $r_{CN}\leq0.42$ nm) in simulations of 1M salt systems. Darker lines show RDFs during the entire MD simulation of 1M salt systems.}
  \label{sfig:a_m_rdfs_bounded_CL_nz}
\end{figure}

\end{widetext} 

\end{document}